\ifx\texsis\undefined \input mtexsis.tex \fi
\texsis
\input LAT96.txs
\input epsf             



\headline={\IDline}\advance\headlineoffset by 0.5cm
\def\IDline{%
   \ifnum\pageno=1 {\hfill
     \vbox to 0pt{\hbox{UM-TH-96-11}\vskip 2pt\hbox{hep-lat/9608111}\vss}}%
   \else {\hfill}\fi}



\def\th{\hbox{-{th}}}
\def\B{\beta}


\titlepage
\title
Why a Particle Physicist is Interested in DNA Branch Migration
\endtitle
\author
Eric Myers${}^a$ and Michael F. Bruist${}^b$
${}^a$Department of Physics, Randall Laboratory, University of Michigan,
Ann Arbor, MI 48109-1120
\smallskip
${}^b$Department of Chemistry, Vassar College, Poughkeepsie, NY 12601
\endauthor

\abstract
We describe an explicitly discrete model of the process of DNA branch
migration.
The model matches the existing data well, but we find that branch
migration along long strands of DNA ($N \simge 40$~bp) is also well
modeled by continuum diffusion.
The discrete model is still useful for guiding future experiments.
\endabstract
\endtitlepage


Lattice methods have proved to be very useful for studying particle
interactions.  As reported at this meeting, lattice gauge theory
calculations now seem to be able to predict the mass of the
``glueball,'' and are leading to a better understanding of both chiral
symmetry breaking and quark confinement.  It is implicit in such
calculations (and sometimes explicit) that one is to take the continuum
limit, but one can, of course, apply lattice field theory methods to
systems which are inherently discrete, without taking the continuum
limit.  DNA branch migration is a simple example.

\bottomfigure{Fig1}
\forceright
\centerline{\epsfxsize=\colwidth\epsfbox[0 15 370 280]{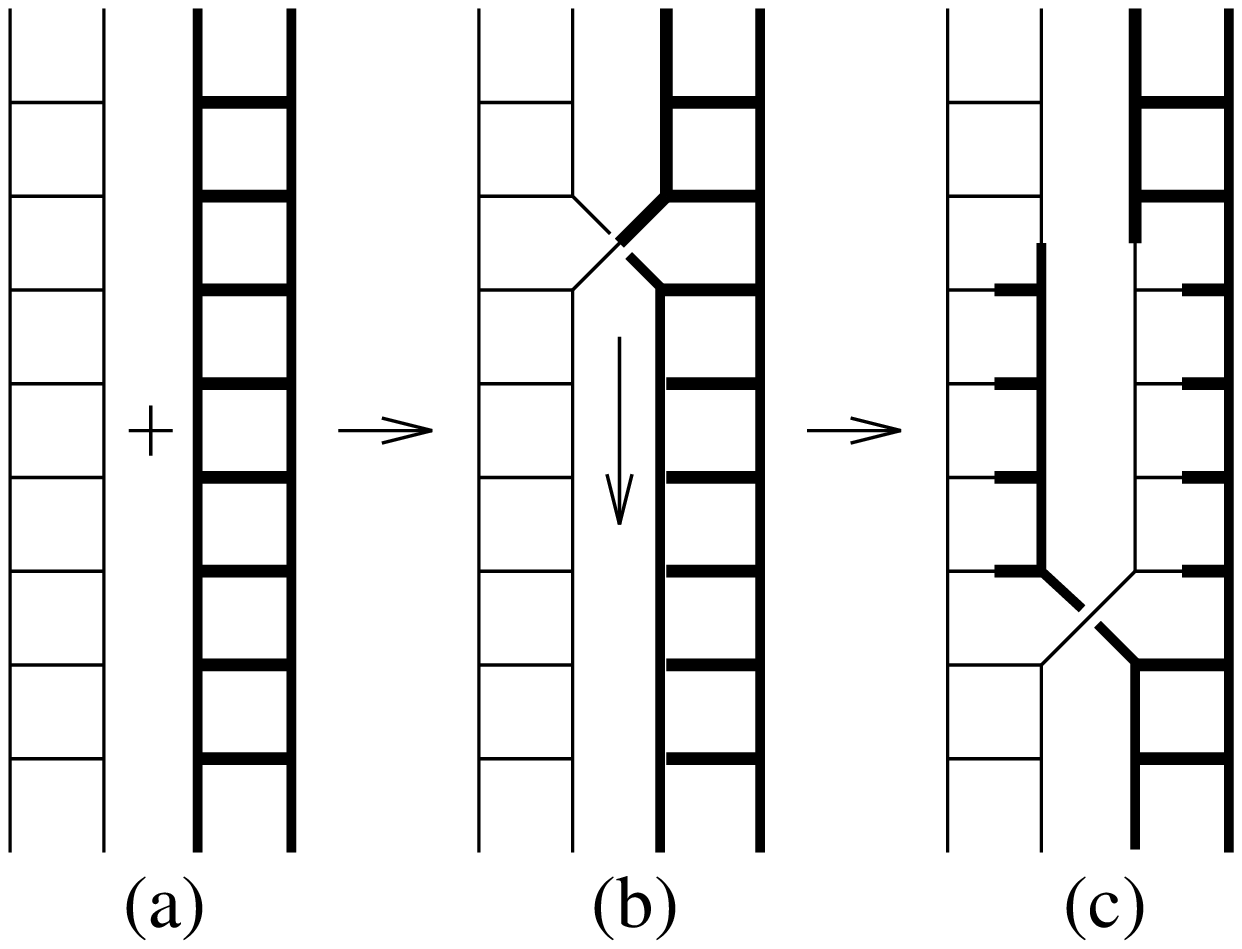}}
\caption{Schematic of single strand exchange in DNA to form the Holliday
structure (b), followed by branch migration (c).}
\endfigure

DNA (deoxyribonucleic acid) transmits genetic information when the
molecule is replicated and transmitted to daughter cells.  Rare errors
in replication give rise to mutations.  (The adaptability of organisms
with these mutations gives rise to evolution.)  Mutations do not stay in
the DNA segment (chromosome) in which they arose; they can transfer to
other chromosomes by {\sl recombination}.  In recombination, two DNA
segments encoding the same genes effectively break at homologous points,
and then rejoin or recombine by fusing partner DNA segments.  When
viewed in more detail, one finds an intermediate state (a "Holliday
structure") in which two DNAs are joined by the exchange of only one
strand from each duplex\reference{Holliday, 1964} R. Holliday,
\journal Genetic Research;5,282-304(1964)
\endreference\relax
(\Fig{Fig1}b).  The branch point which joins the two helices is mobile
because of the sequence similarity of the homologous DNAs (\Fig{Fig1}c).
A better understanding of branch migration will lead to a better
understanding of DNA recombination and the propagation of genetic
variation which drives evolution.

Laboratory experiments can now replicate DNA branch migration through
short defined segments of DNA with controlled initial
conditions.\reference{Bruist, 1996} A.W.~Kirby, M.N.~Gaskin,
M.A.~Antezana, S.J.~Goodman, E.~Myers, and M.F.~Bruist,
submitted to {\sl J.~Molecular Biology}
\endreference
\reference{Panyutin, 1995}
I.G. Panyutin,  I. Biswas, and P. Hsieh,
\journal EMBO J.;14,1819-1826(1995)
\endreference
Our goal is to develop a model of these experiments which preserves the
discrete nature of DNA. We focus on the motion of the branch point and
ignore the helical structure of DNA.  (A related model has been
described by Fujitani and Kobayashi.\reference{Fujitani, 1995}
Y. Fujitani and I. Kobayashi,
\journal Physical Review E;52,6607-6622(1995)
\endreference)
This can be thought of as a series of linked chemical reactions in a
``compartment model.''\reference{Bentley and Cooke} D.L. Bentley and K.
L. Cooke, \booktitle{Linear Algebra with Differential Equations}, (Holt,
Rinehart and Winston, 1973)
\endreference
Imagine that all of the molecules with branch points in a particular
position are kept in one
compartment, and that when the branch point on a given molecule moves to
a neighboring position that molecule is removed from its compartment and
added to the appropriate neighboring compartment.  If there are $N$
possible positions for the branch then there will be $N+1$ compartments,
the last being for molecules which have separated.  Ignoring this last
compartment, let the $n\th$ component of the vector $\vec C(t)$
represent the concentration of molecules having the branch in the $n\th$
position at time $t$, and let $\B$ represent the rate of transport in
the forward or backward directions (assuming these are equal---it is
easy to generalize to the case where they are not). The dynamics of this
system is then described by a set of $N$ coupled first order linear
differential equations, which can be written in matrix form as
$$\EQNdisplaylines{
{d \vec C \over dt} = \beta a^2 \bar M \vec C   \,,
\hfill\EQN 7}$$
where the $N\times N$ matrix $\bar M$ is
$$\EQNdisplaylines{
\bar M = {1 \over a^2}
\pmatrix{
 -1 & ~1     &        &      &    \cr
 ~1 & -2     & 1      &      &    \cr
    & \ddots & \ddots &\ddots&    \cr
    &        & 1      & -2   & ~1  \cr
    &        &        & ~1   & -2  \cr} \,.
\hfill\EQN 8}$$
All blank entries are zero, and we have introduced $a$ as the spacing
between successive positions along the DNA.

The similarity of \Eq{8} to the lattice second derivative is obvious,
and so it should come as no surprise that the continuum limit of this
model is the diffusion equation.  However,
we seek solutions to \Eq{7} before the continuum
limit is taken.  The procedure for finding these solutions is
similar to the continuum case.  First, one can remove a time-dependent
factor by writing $\vec C(t) = e^{-t/\tau} \vec v$, where $\vec v$ is an
eigenvector of $\bar M$.  If the $n\th$ component
of $\vec v$ is written as $v(n)$, then the eigenvectors of $\bar M$ have
the form
$$\EQNdisplaylines{
v(n) = \cos(kna + \phi) \,,
\hfill\EQN 10}$$
where $n=0,1,\ldots,N-1$, and where $k$ and $\phi$ are yet to be
determined constants. The eigenvalues are
$$\EQNdisplaylines{
\lambda = {-4 \over a^{2}} \sin^2({ka \over 2})         \,,
\hfill\EQN 11}$$
from which it follows that
$$\EQNdisplaylines{
\tau = {- 1 \over \B a^{2} \lambda}  \,.
\hfill\EQN 13}$$

In the continuum case the undetermined constants $k$ and $ \phi$ would
be determined by boundary conditions, but there are no ``boundaries'' to
a matrix.
Nevertheless, it is useful to imagine that there are extra compartments
at either end of the system.
\Eq{10} is only an eigenvector because it
fits the +1,~-2,~+1 differencing pattern along the diagonals of $\bar
M$.
This pattern must be maintained for the eigenvalue equation to be valid.
To do so at the upper left corner of $\bar M$ we imagine an extra point
at $n=-1$ such that $v(-1) = v(0)$, which forces $\phi = \half ka$.
This says that the lattice derivative is zero, so in analogy to the
continuum case we call this a ``lattice Neumann'' boundary condition.
To maintain the differencing pattern at the lower right corner of $\bar
M$ we imagine an extra point at $n=N$, such that $v(N) = 0$ (this is the
final compartment that we ignored earlier).
This ``lattice Dirichlet'' boundary condition forces the unknown
constant $k$ to be
$$\EQNdisplaylines{
k = {(2m + 1) \over (2N + 1)} \, {\pi \over a}  \,,
\hfill\EQN 19}$$
where $m=0,1,2 \ldots N-1$.
There are exactly $N$ distinct eigenvalues, which we
will label $\vec v_m$.

\bottomfigure{Fig2}
\forceright \epsfxsize=\colwidth
\centerline{\epsfbox[50 90 550 450]{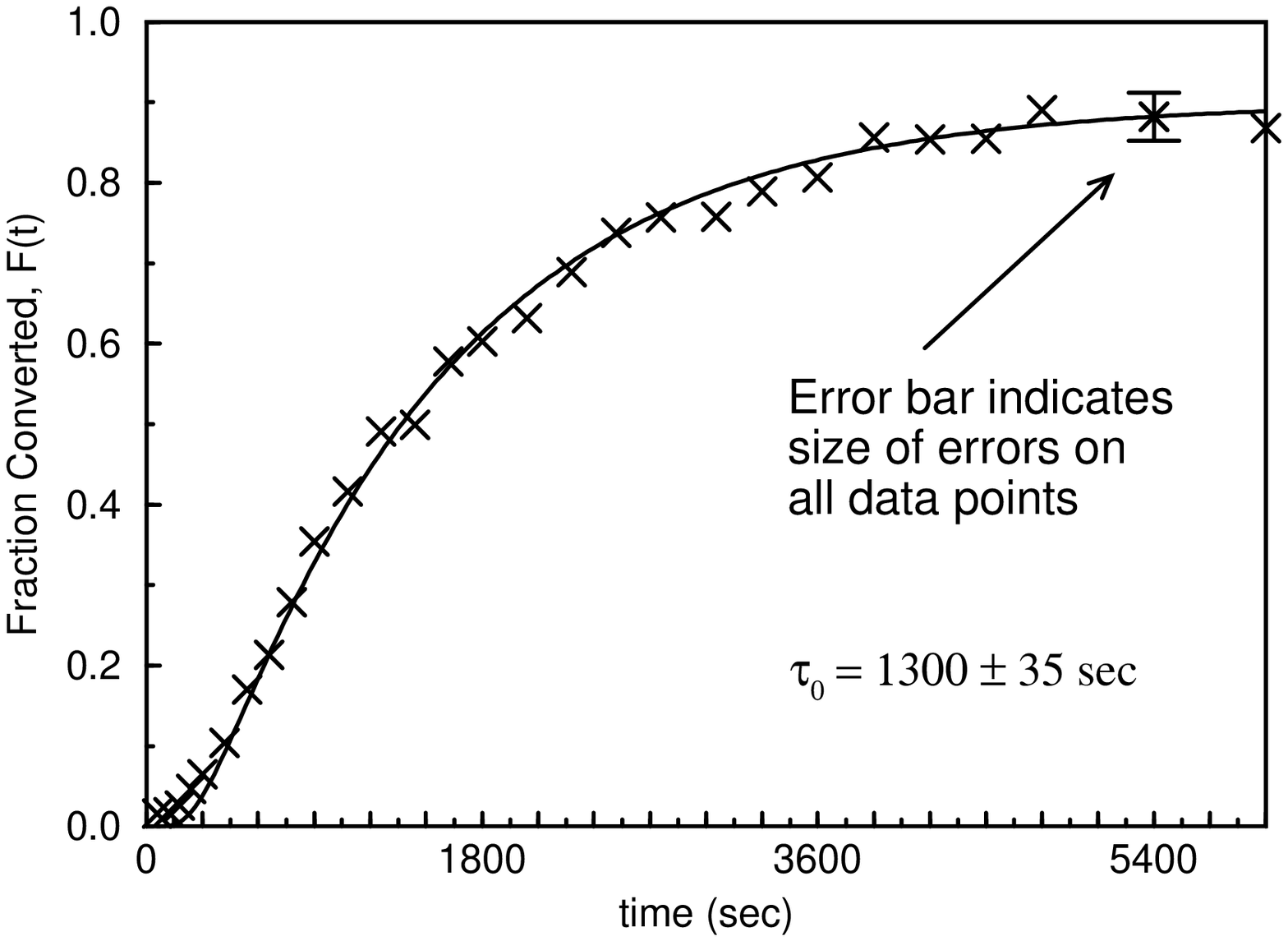}}
\caption{Data from an experiment measuring the fraction of branched DNA
converted to separated final product, along with the best fit curve from
the model, \Eq{28}, for $N=40$.}
\endfigure

The most general solution of \Eq{7} is a linear combination of all
of the solutions,
$$\EQNdisplaylines{
\vec C(t) = \sum_{m=0}^{N-1}
         a_{m} \, e^{-t/\tau_m} \, \vec v_m         \,,
\hfill\EQN 23}$$
where the $a_m$ are constants, and where $\tau_m$ is related to the
$m\th$ eigenvalue via \Eq{13}.
The values of the $a_m$ are determined from the initial conditions,
using the fact that the different eigenvectors $\vec v_m$ are
orthogonal.
In the experiments we are modeling,\cite{Bruist, 1996} the initial
conditions are such that all molecules begin with the branch point at
one position, adjacent to a reflective barrier created by a divergence
in the DNA sequences.
These initial conditions are represented by $\vec C(0) = C_0
\delta_{n,0}$.  Substituting this into \Eq{23}, taking the dot product
with $\vec v_{m^\prime}$ and using the orthogonality of the $\vec v_m$
yields
$$\EQNdisplaylines{
a_m = {4 C_0 \over 2N + 1}
        \cos({\pi \over 2} \, {(2m+1) \over (2N+1)} )
\hfill\EQN 26}$$

The experiments we model only measure the concentration of separated
molecules at a given time, not the number of molecules having the branch
point in a given position.  To compute this observable we take the
initial concentration, $C_0$, and subtract from it the sum over all of
the branched molecules left in the system.  This gives

$$\EQNdisplaylines{
F_N(t)
     = C_0  \Biggl\lbrack 1 - {2 \over 2N+1} \sum_{m=0}^{N-1}
        (-1)^m  e^{-t/\tau_m} \hfill \cr
\hfill{}\times
        {\cos^2({\pi \over 2} \, {2m+1 \over 2N+1}) \over
         \sin  ({\pi \over 2} \, {2m+1 \over 2N+1}) } \Biggr\rbrack
.\qquad\EQN 28\cr}
$$
Aside from the number of steps, $N$, there are only two free parameters
in this expression, $C_0$ and $\B$ (which is related to the $\tau_m$ by
\Eq{13}).

\figure{Fig3}
\forceright \epsfxsize=\colwidth
\centerline{\epsfbox[20 90 550 450]{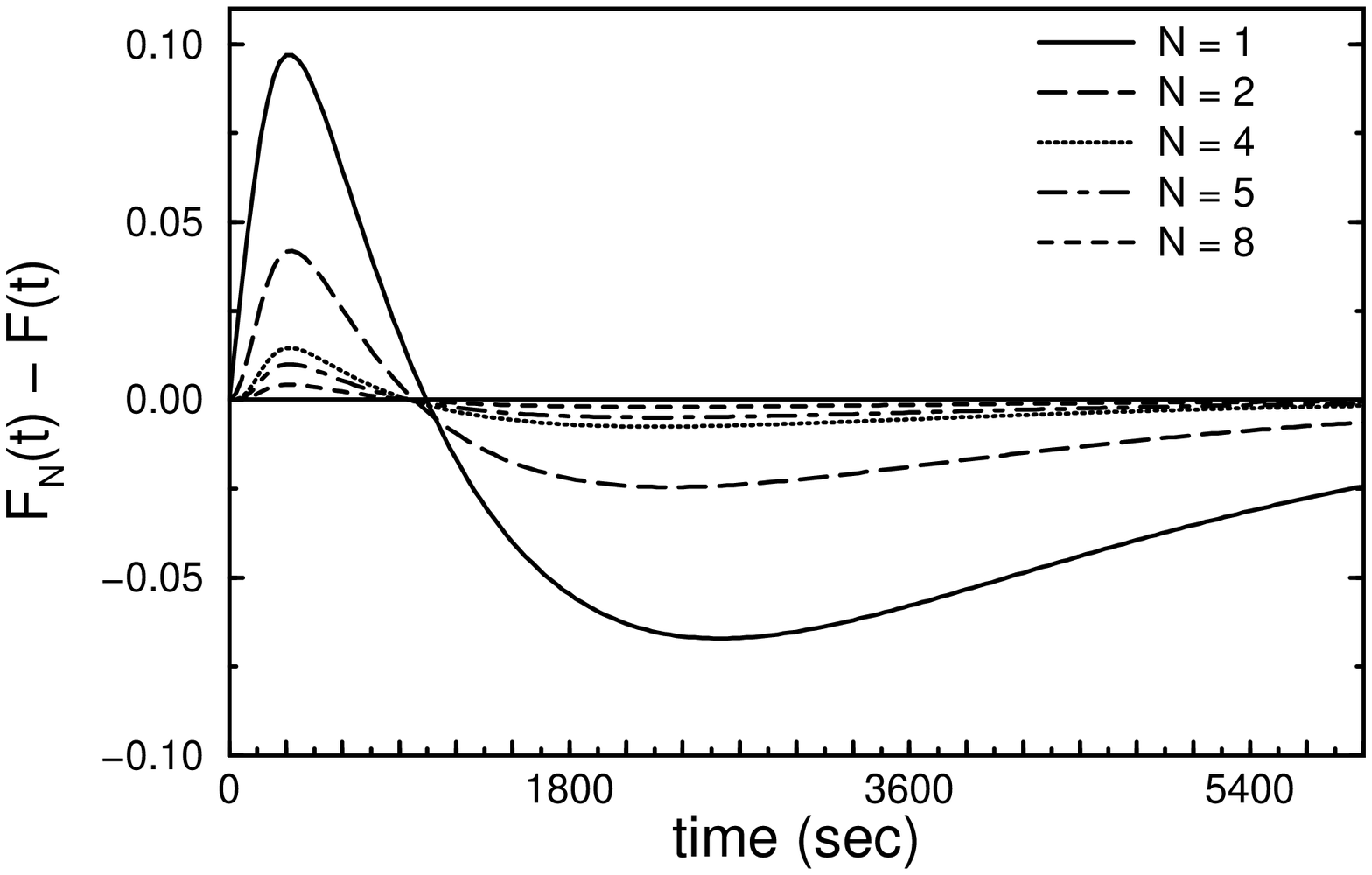}}
\caption{Difference between the best fit discrete and continuum measures
of the final product.}
\endfigure

\Fig{Fig2} shows measurements of the concentration of final product
(separated molecules) from an experiment using a DNA junction in which
the branch point must migrate 40~base pairs (bp) in one direction before
coming to the end of the strands.
The errors include systematic errors, not just statistical variations.
The curve in the figure is a non-linear least-squares fit to \Eq{28}
with $N=40$.
The fit is quite good.
One notable characteristic of branch migration is the ``lag'' at the
beginning of the experiment---no material is produced as final product
for several minutes after the experiment has begun.
The theoretical curve reproduces this behavior well.

Is it possible to see the discrete nature of the motion of the branch point?
In particular, can we tell the number of base pairs traversed in a
single step of branch migration?
Unfortunately, the answer is no.  The data in
\Fig{Fig1} are also fit well by \Eq{28} with $N$ anywhere between 4 and 40,
as well as by the analogue of \Eq{28} obtained from
continuum diffusion.  \Fig{Fig3} shows the difference between
$F_N(t)$ from \Eq{28} and $F(t)$ obtained from continuum diffusion
(assuming the same $C_0$).
Except for very short molecules,
the differences are smaller than the size of present experimental
errors.
Nevertheless, \Fig{Fig3} shows that the discrete nature of branch
migration will be most noticeable within times less than $\tau_0$.
This is useful knowledge for designing future experiments.
We therefore expect that our discrete model will lead to a better
understanding of the underlying dynamics of branch migration.
\nobreak
\nosechead{REFERENCES}%
\ListReferences
\bye